\documentclass[runningheads]{llncs}
\usepackage[T1]{fontenc}
\usepackage{graphicx}
\usepackage{amsmath}
\usepackage{algorithm}
\usepackage{algpseudocode}
\usepackage{balance}
\usepackage{color}
\usepackage{multicol}
\usepackage{multirow}
\begin{document}
\title{Modality-Agnostic Learning for Medical Image Segmentation Using Multi-modality Self-distillation}


%
\titlerunning{Modality-Agnostic Learning for Medical Image Segmentation}

\author{}
\institute{}

\author{Qisheng He\inst{1} \and
Nicholas Summerfield\inst{2,3} \and
Ming Dong\inst{1} \and
Carri Glide-Hurst\inst{2,3}
}
%
%
\institute{
Wayne State University \\
Department of Computer Science \\
5057 Woodward Ave, Detroit, MI 48202 
\\ \and
University of Wisconsin-Madison\\
Department of Human Oncology\\
Department of Medical Physics\\
600 Highland Ave, Madison, WI 53792\\
}
\maketitle              

\begin{abstract}
Medical image segmentation of tumors and organs at risk is a time-consuming yet critical process in the clinic that utilizes multi-modality imaging (e.g, different acquisitions, data types, and sequences) to increase segmentation precision. In this paper, we propose a novel framework, Modality-Agnostic learning through Multi-modality Self-dist-illation (MAG-MS), to investigate the impact of input modalities on medical image segmentation. MAG-MS distills knowledge from the fusion of multiple modalities and applies it to enhance representation learning for individual modalities. Thus, it provides a versatile and efficient approach to handle limited modalities during testing. Our extensive experiments on benchmark datasets demonstrate the high efficiency of MAG-MS and its superior segmentation performance than current state-of-the-art methods. Furthermore, using MAG-MS, we provide valuable insight and guidance on selecting input modalities for medical image segmentation tasks.
\vspace{-0.1in}
\end{abstract}
\section{Introduction}
Medical image segmentation is a time consuming task performed by expert physicians. Deep learning is thus involved for more accurate and efficient segmentation. In medical imaging, there are usually multiple modalities/sequences (note that for simplicity, the term ``modality'' here also applies to varied acquisition sequences as well) that focused on the same target, where each modality contains feature-rich and biological information \cite{brats}. Transfer learning \cite{transfer-learning} is one such method to integrate multi-modality data into segmentation by leveraging the information obtained from a pre-trained model on one source modality and applying it to another target modality. Another technique, multi-modality fusion such as input-level\cite{unetr}, feature-level \cite{fullyConnectedCRF}, and decision-level fusion \cite{fullyConvNets}, employs information from different imaging modalities such as MRI, CT, and PET scans to provide a comprehensive understanding of tumor extent and improve segmentation accuracy \cite{multimodalMedicalImg}. 

A recent study has shown the importance of input modality selection on the performance of deep learning models for head and neck tumor segmentation. In \cite{evaluation}, a deep neural network is repeatedly trained and evaluated on different input combinations selected from seven imaging modalities. However, such a brute-force approach across all model combinations requires a vast amount of computational resources, training time, and may be limited by the model expecting the same number and image modalities during both training and testing. That is, each network is trained with a specific combination of modalities and, unless another technique such as transfer learning is used, may only evaluate the same combination of datasets. This characteristic greatly limits a network’s application in the clinic where datasets may be incomplete or vary from clinic to clinic \cite{MR-only-review}.

Although multi-modality training is possible, clinical translation is challenged by the integration of complex datasets for segmentation tasks and the limited availability of all image types, acquired with the same parameters and conditions, for any given patient.  Thus, when training a model, modality-agnostic (MAG) learning is a more efficient and effective approach. Different from existing methods, MAG learning trains a single model based on all available modalities but remain input-agnostic, allowing a single model to produce accurate delineations on any number of different modality combinations, even a single input dataset \cite{modalityVariantRec}. Current MAG learning methods handle partial available modalities during testing by using mean fusion with modality dropout training \cite{hemis}, or by filling missing modalities with all zeros or means \cite{handlingMissingSequences}.

In this paper, we propose a novel framework for MAG learning through Multi-modality Self-distillation (MAG-MS) and apply it to investigate the effects of different input combinations of modalities in medical image segmentation. Recently, the self-distillation technique \cite{selfDistillation} has been proposed for knowledge transfer \cite{kdOnMultiModalAndMonoModal} within the same model by allowing a network to distill its own knowledge through the use of its intermediate representations and outputs. The network in MAG-MS consists of unique modality-specific encoders and a shared common decoder that extracts modality-specific features while leveraging complementary information from multiple modalities. Modality-specific features are fused through averaging and fed into the shared decoder, whose output is used as the teacher to guide individual modalities (students) through self-distillation. Importantly, during testing, features of unavailable modalities are dropped so that the model can handle different input modality combinations without the need for retraining. The main contributions of this work can be summarized as follows:

1. MAG-MS provides a versatile and efficient approach to handling missing modalities during testing. Its unique modality-agnostic structure requires a \textbf{single} training to perform segmentation on different input combinations of imaging modalities, training in a fraction of the time.

2. MAG-MS distills knowledge from the fusion of multiple modalities and applies it to enhance representation learning for individual modalities. We prove mathematically that training through multi-modality self-distillation improves model testing performance on partially available input modalities.

3. Our extensive experiments on benchmark datasets showed that MAG-MS outperforms existing MAG-learning methods by a \textbf{significant margin}. Additionally, using our MAG-MS framework, we provided valuable insights and guidance on the most powerful input modality choices for medical image segmentation tasks, tailored to specific anatomical requirements for use cases.

\section{MAG Learning with Multi-modality Self-distillation}
\begin{figure}[t]
    \centering
    \vspace{-0.15in}
    \includegraphics[width=0.9\textwidth]{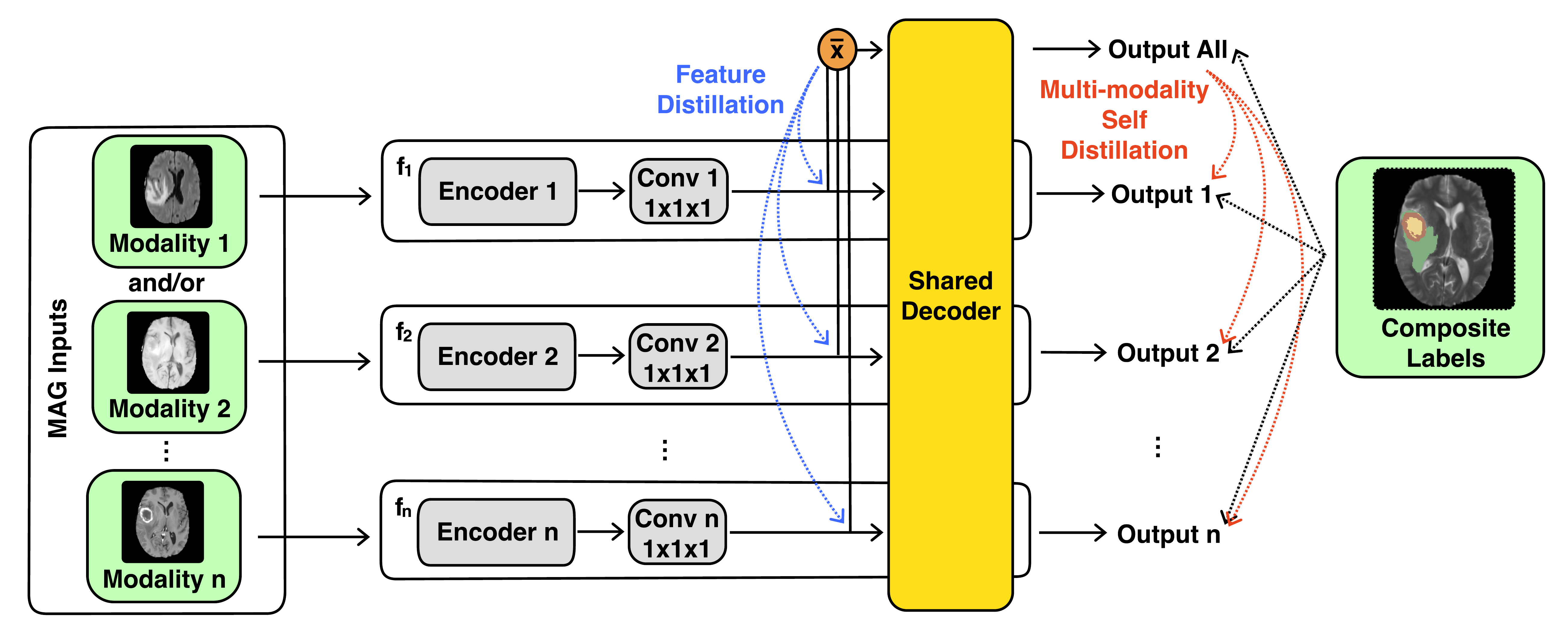}
    \vspace{-0.15in}
    \caption{MAG-MS architecture illustration. During training, all-modalities are available a single set of composite ground-truth labels, and the all-modalities features and outputs are used to distillate single-modality ones. During testing, paths for unavailable modalities are skipped highlighting the flexibility of the model.}
    \label{fig.structure}
\end{figure}
\setlength{\textfloatsep}{12pt}

\subsection{Feature Fusion in MAG Learning}
The overall architecture of our framework is shown in Fig. \ref{fig.structure}. To handle limited/missing input modalities (MAG inputs) during testing, we designed our MAG-MS model with unique modality-specific encoders and a shared decoder. When multiple modalities are available in training, feature-maps of different modalities are fused before passing them to the shared decoder. Specifically, to keep our model modality agnostic, we use a $1\times1\times1$ point-wise convolutional layer to extract a linear combination of features from each modality \cite{mobilenets} and then calculate the mean of these features instead of concatenating them \cite{multistreamFCN}.

Let's assume that there are $M$ modalities in total, and the modality-specific encoder $enc_i$ is used for an input $x_i$ for modality $i \in M$. So, the bottleneck feature of modality $i$ is obtained as $\mathbf{F}_i = enc_i(x_i)$. In MAG-MS, we used a shared decoder $dec$ across modalities since all modalities are targeted to the segmentation task through the same set of labels (composite labels). The output of modality $i$ can hence be denoted as $f_i(x_i) = dec(enc_i(x_i))$. As shown in Fig. \ref{fig.structure}, features from different modalities are fused through averaging when multiple modalities are available. The fused features $\tilde{\mathbf{F}} = \sum^M_i enc_i(x_i) / \|M\|$ are then used to calculate the fused outputs $\tilde{f}(x) = dec(\tilde{\mathbf{F}})$, where $x$ denotes the multi-modality input. During testing, the fused features are computed from available modalities in a subset $S \subset M$ as $\tilde{\mathbf{F}}_S = \sum^S_i enc_i(x_i) / \|S\|$ whereas unavailable inputs are dropped.

\subsection{Multi-Modality Self-Distillation for MAG Learning}
In the training of MAG-MS, we self-distillate \cite{selfDistillation} both features and outputs of each modality ($\mathbf{F}_i$ and $f_i$) as students using the fused ones ($\tilde{\mathbf{F}}$ and $\tilde{f}$) as the teacher. In a modality-agnostic network, the network has already been divided into different modality sections and thus self-distillation can be directly performed. There are three kinds of losses involved in multi-modality self-distillation: the dice cross entropy loss, the KL-Divergence loss, and the feature-wise L2 loss.

First, the dice cross entropy loss is a composite label-guided combined loss of dice and cross entropy losses, where the cross entropy is mainly used in classification, and the dice loss solves the data imbalance problem associated with image segmentation tasks \cite{diceLoss}. Such a combination is widely used in medical image segmentation. In MAG-MS, the dice cross entropy loss $DC(y; f_i(x_i))$ is directly computed from the composite labels $y$ of the training dataset and the output of $i^{th}$ modality. Second, the KL-Divergence loss is widely used in knowledge distillation for image classification \cite{knowledgeDistillation}, where a student tries to mimic the teacher via the distance of soft predictions that are obtained by temperature-calibrated softmax of the logits. In MAG-MS, the pixel-wise KL-Divergence loss $D_{KL}(\tilde{f}(x); f_i(x_i))$  \cite{paSeg} is calculated by using the output $f_i(x_i)$ of the $i^{th}$ modality as the student and the fused all-modalities output $\tilde{f}(x)$ as the teacher. Finally, feature-wise L2 loss computes the L2 distance between the features from a single-modality $\mathbf{F}_i$ and the fused features from all-modalities $\tilde{\mathbf{F}}$.

By summing these three losses, the combined self-distillation loss for each modality $i$ is given as:
\vspace{-0.05in}
\begin{equation}
l_i(y; \tilde{f}(x); f_i(x_i)) = DC(y; f_i(x_i)) + \lambda D_{KL}(\tilde{f}(x); f_i(x_i)) + \gamma \|\mathbf{F}_i - \mathbf{\tilde{F}}\|^2
\vspace{-0.05in}
\end{equation}
where $\lambda$ and $\gamma$ are the balance factors for the KL-divergence loss and the L2 loss, respectively.
 
\subsection{Modality-Agnostic Loss in MAG Learning}
In MAG-MS, we use a modality-agnostic loss (MAG loss) to optimize the network. Particularly, the input features of each single-modality $i \in M$ and the fused features are forward passed one-by-one in a single training iteration to generate their corresponding outputs. The MAG loss is then computed by summing the losses incurred by each individual modality and the fused all-modalities before being used to update the network weights through back-propagation:
\vspace{-0.12in}
\begin{equation}
    \label{eq.magLoss}
    l(y; \tilde{f}(x); f_i(x_i)) = \tilde{l}(y; \tilde{f}(x)) + \sum^M_i l_i(y; \tilde{f}(x); f_i(x_i))
    \vspace{-0.12in}
\end{equation}
where $\tilde{l}(y; \tilde{f}(x)) = DC(y; \tilde{f}(x))$ denotes the dice cross entropy loss associated with the fused all-modalities output. The detailed training algorithm is shown in Algorithm \ref{alg.magms}.

\begin{algorithm}[t]
    \caption{MAG-MS Training Procedure}
    \label{alg.magms}
    \begin{algorithmic}[1]
        \Function{Train}{$Dataset$, $M$, $iterations$}
        \For{$t$ in $iterations$} \Comment{Training loop}
            \State $x, y \leftarrow Dataset[t]$ \Comment{Get data}
            \For{$i$ in $M$} \Comment{Forward modalities}
                \State $\mathbf{F}_i \leftarrow enc_i(x_i)$ \Comment{Fetch features for modality $i$}
                \State $f_i(x_i) \leftarrow dec(\mathbf{F}_i)$ \Comment{Fetch output for modality $i$}
            \EndFor
            \State $\tilde{\mathbf{F}} \leftarrow$ fuse($\mathbf{F}_i$) \Comment{Fetch fused features} 
            \State $\tilde{f}(x) \leftarrow dec(\tilde{\mathbf{F}})$ \Comment{Fetch fused outputs} 
            \State $loss \leftarrow$ $l$($y$; $\tilde{f}(x)$; $f_i(x_i)$) \Comment{Calculate modality-agnostic loss}
            \State backPropagate($loss$)
        \EndFor
        \EndFunction
    \end{algorithmic}
\end{algorithm}
\setlength{\textfloatsep}{6pt}

\subsection{Theoretical Basis}
Let $\Theta_{dec}$ be the shared parameters of the network decoder. During training with all $M$ modalities, the prediction of the probabilistic model $P(\tilde{f}(x)|x)$ is maximized to obtain $\theta_M \in \Theta_{dec}$. During testing, the prediction is obtained by maximizing the probabilistic model $P(f_i(x_i)|\{x_i; i \in S\})$. In Theorem 1 of \cite{foundationsOfMCL}, through the universal approximation capabilities of neural networks, it was shown that if the domain of a learning process is both strictly multi-modal and interdependent (see \cite{foundationsOfMCL} for definitions), during training, each modality in $S$ can learn a mapping to features in other modalities in $M$ and hence improve its performance in isolation during testing. This provides the theoretical basis for MAG-learning. Now, we show that multi-modality self-distillation in our framework can further improve the testing performance on partially available input modalities $S$.

\textbf{Theorem 1.} \textit{There exists $\theta_M$ and $\theta'_M$ such that the following inequality holds through multi-modality self-distillation:}
\begin{equation}
\label{eq.theorem1}
\exists \theta_M, \theta'_M \in \Theta_{dec}; H(f_i(x_i)|\{x_i; i \in S\}; \theta'_M) < H(f_i(x_i)|\{x_i; i \in S\}; \theta_M)
\end{equation}
\textit{where $H(\cdot|\cdot; \theta_M)$ and $H(\cdot|\cdot; \theta'_M)$ denote the conditional entropy of the model output given testing inputs using $\theta_M$ and $\theta'_M$, respectively; $\theta_M$ is learned in the shared decoder without self-distillation, and $\theta'_M$ is learned with multi-modality self-distillation.} The proof is given in the Appendix.

\section{Experimental Details and Results}
We empirically study the performance of MAG-MS on two medical image segmentation datasets: 2017 infant brain MRI segmentation challenge (iSEG2017) \cite{iseg2017} and Medical Segmentation Decathlon Task 01 (MSD) \cite{msd}. We used UNETR \cite{unetr}, a U-shaped network with a vision transformer as its encoder to directly connect to its decoder via skip connections at different resolutions, as our backbone model. We set both $\gamma$ and $\lambda$ as 1 in the multi-modality self-distillation for all experiments. The same number of epochs and learning rates were implemented during training \cite{hyperDenseNet,unetr}. To compare against the state of the art methods, we implemented and trained existing modality-agnostic models including HeMIS \cite{hemis} and UNETR with missing modalities filled by zeros and means (baselines) \cite{handlingMissingSequences} with the same settings for a fair comparison. Quantitative evaluations between predictions and ground truth segmentation were performed using the Dice similarity coefficient and Hausdorff Distance 95 (HD95) \cite{diceOverlap,nnUNet}, where results are reported with mean $\pm$ variance. All experiments were implemented with a PyTorch \cite{pytorch} and Monai \cite{monai} framework and run on a machine with two NVIDIA RTX A6000 GPUs. The source code of this work will be made publicly available after the double-blind review period of MICCAI 23. 

\subsection{Infant Brain MRI Segmentation (iSEG2017)}
We first performed an ablation study on the iSEG2017 challenge dataset to demonstrate the impact of multi-modality self-distillation on MAG learning. The dataset contains 10 volumes of 6-month infant brain tissues from MR scans with two modalities, T1-weighted ($T_1$) and T2-weighted ($T_2$). Following \cite{hyperDenseNet}, the dataset was separated into training (6), validation (1), and testing (3) sets. 

\begin{table}[t]
    \centering
    \includegraphics[width=0.8\textwidth]{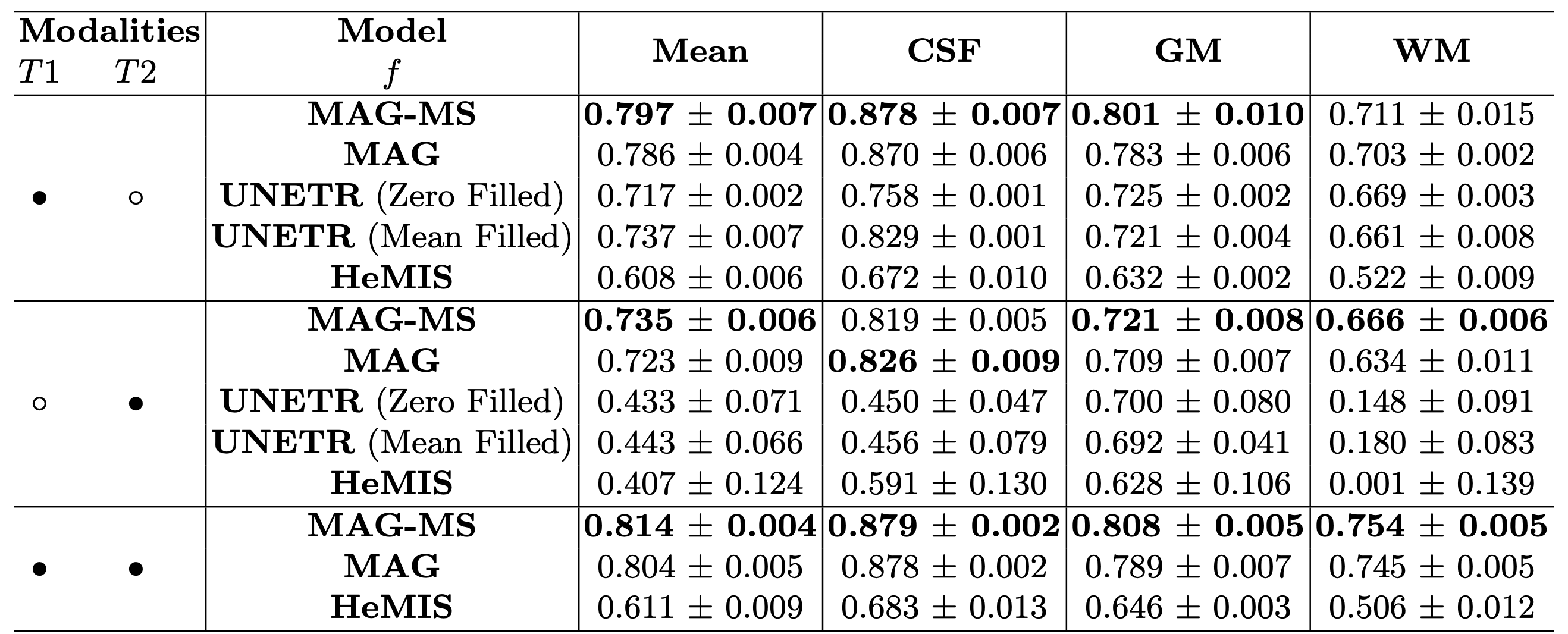}
    \caption{Dice scores on iSEG dataset obtained by MAG-MS, MAG Learning without self-distillation (MAG), UNETR with missing modalities filled by zeros or mean, and HeMIS for Cerebrospinal Fluid (CSF), Grey Matter (GM), and White Matter (WM). The highest scores are \textbf{bolded}, and a $\bullet$ indicates that the corresponding modalities are available during testing. The same applies to Tables \ref{table.msdDice} and \ref{table.msdUNETR}.}
    \label{table.iseg}
\end{table}
\setlength{\textfloatsep}{6pt}

Table \ref{table.iseg} displays the results on iSEG2017 obtained by MAG-MS, compared to other modality-agnostic methods. Even without multi-modality self-distillation (MAG), our model already outperformed other methods on all MAG inputs, which achieved a mean Dice score of 0.786 on $T_1$, 0.723 on $T_2$, and 0.804 on fused all-modalities, respectively. Meanwhile, the multi-modality self-distillation improved MAG for more than 1\% on all MAG inputs, which achieved a mean Dice score of 0.797 on $T_1$, 0.735 on $T_2$, and 0.814 on fused all-modalities, respectively. The results also reveal that $T_1$ provided more discriminative information for brain segmentation. This is evidenced by the performance of MAG-MS, which achieves higher Dice scores on the $T_1$ modality (0.797) than on $T_2$ (0.735). 

\subsection{Medical Segmentation Decathlon Task 01 (MSD)}
Our MAG study was then expanded to the MSD dataset which comprises 484 brain tumor volumes, each featuring 4 modalities: $T_1$, $T_1c$ ($T_1$ with Gadolinium contrast enhancement), $T_2$, and $F$ (T2-W Fluid Attenuated Inversion Recovery). Following \cite{unetr}, we partitioned the dataset into training (388), validation (72), and testing (24) sets.

\begin{table}[t]
    \centering
    \includegraphics[width=0.7\textwidth]{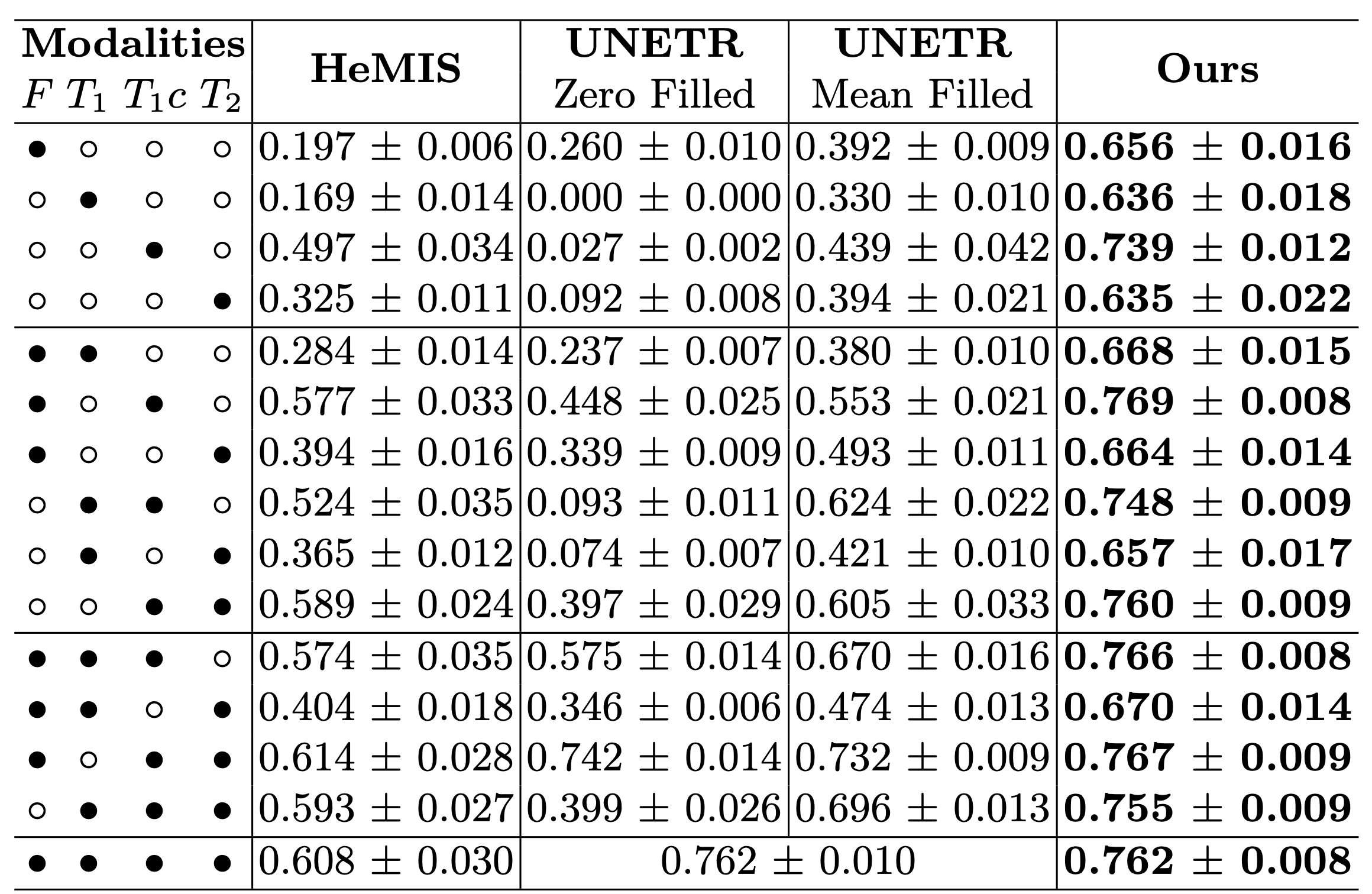}
    \caption{Mean Dice scores on varied MAG inputs of the MSD dataset, obtained by HeMIS, UNETR (missing modalities filled by zeros or mean), and our MAG-MS model.}
    \label{table.msdDice}
    \vspace{-0.3in}
\end{table}

\begin{figure}[t]
    \centering
    \includegraphics[width=0.9\textwidth]{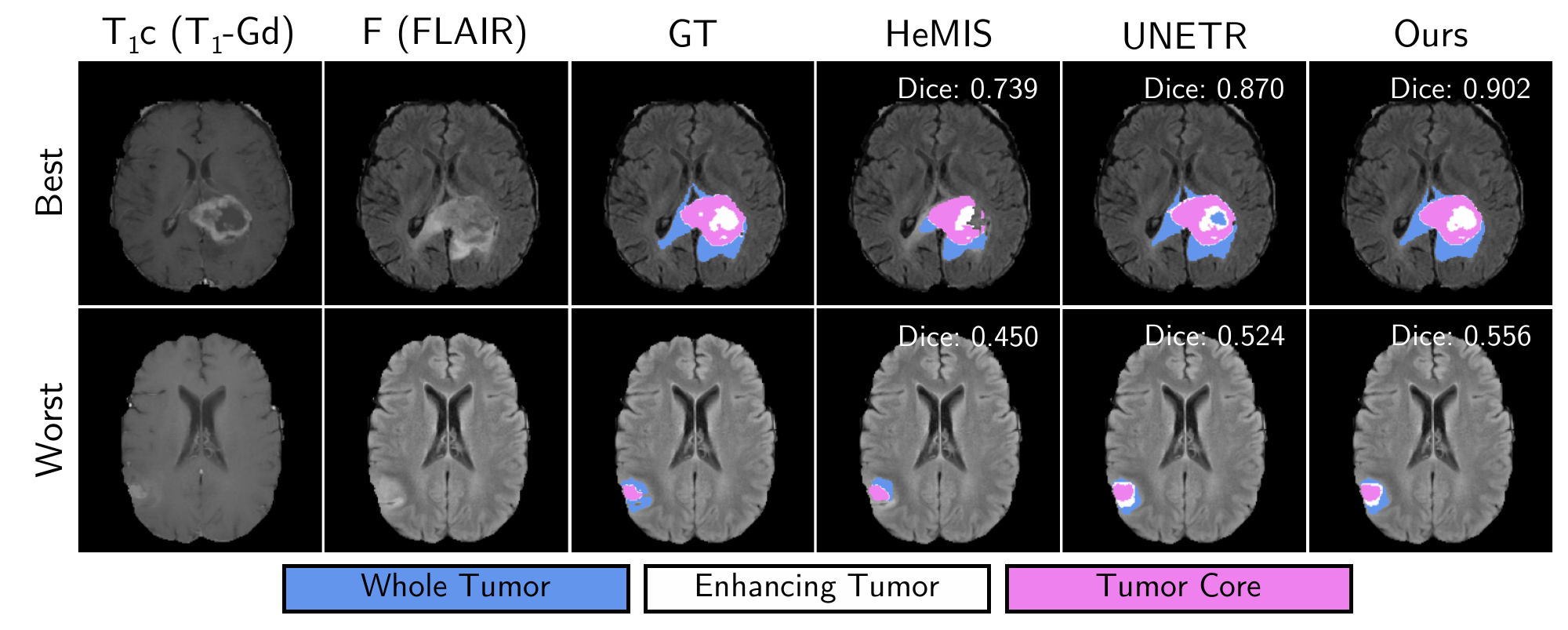}
    \vspace{-0.15in}
    \caption{Best (top) and worst (bottom) prediction masks for HeMIS, UNETR, and our model based on Dice. Predictions were made with all-modality inputs ($F$, $T_1$, $T_1c$, $T_2$). 
    }
    \label{fig.predictions}
\end{figure}
\setlength{\textfloatsep}{6pt}

Table \ref{table.msdDice} shows the results on MSD obtained by MAG-MS, compared to those obtained by other modality-agnostic methods. Again, our method outperforms others with \textbf{a significant margin}, achieving a Dice score of 0.762 when all modalities are available. Here, as may be expected for brain tumor segmentation, $T_1c$ carries the greatest segmentation power, as when it is dropped as an input, there is a significant reduction of about 10\% in Dice score (0.670). Moreover, $T_1c$ achieves a Dice score of 0.739 alone. For primary brain tumors, our results highlight that the optimal input combination was realized for $F$ + $T_1c$ input modalities yielding a Dice score of 0.769. 

Both anatomically and as shown by our findings, $T_1c$ provides excellent visualization of the tumor core (see Appendix for MAG-MS metrics for individual structures) whereas $F$ has great visualization of both the enhancing and whole tumor. These findings are consistent with the literature where clinicians have identified $F$ + $T_1c$ MRI modalities as having the best performance for manual segmentation \cite{BraTS-bestInputs}. 
So, MAG-MS is actually capable of identifying and utilizing the optimal input combination to yield the best possible segmentation results through just a single training. While results are shown for brain segmentation, other extensions of MAG-MS include incorporating functional images and other disease sites, e.g., prostate \cite{MR-only-review} and head and neck \cite{evaluation}.

Best and worst MSD prediction masks for all-modality inputs are shown in Figure \ref{fig.predictions}, directly comparing the different models trained (see Appendix for MAG-MS prediction masks for all other input combinations). In the best case, MAG-MS showed excellent agreement with the ground truth over both HeMIS and UNETR, conserving the enhancing tumor region while better elucidating the complex tumor core and whole tumor regions. The worst prediction results were obtained of a smaller sized lesion on the lateral side of the brain, although MAG-MS still outperformed the other two models and maintained key features.

\subsection{Comparison with individually trained UNETRs}
Table \ref{table.msdUNETR} presents a comparison of our results with each UNETR model trained individually using different input combinations. It is noteworthy that the Dice scores obtained by MAG-MS are very close to those of the UNETR models, while HD95 shows a remarkable improvement on almost all MAG inputs. Overall, 13 out of the 15 models trained by MAG-MS outperform the individually trained UNETR models in terms of Dice scores, and 14 out of 15 are better for HD95.

\begin{table}[t]
    \centering
    \includegraphics[width=0.7\textwidth]{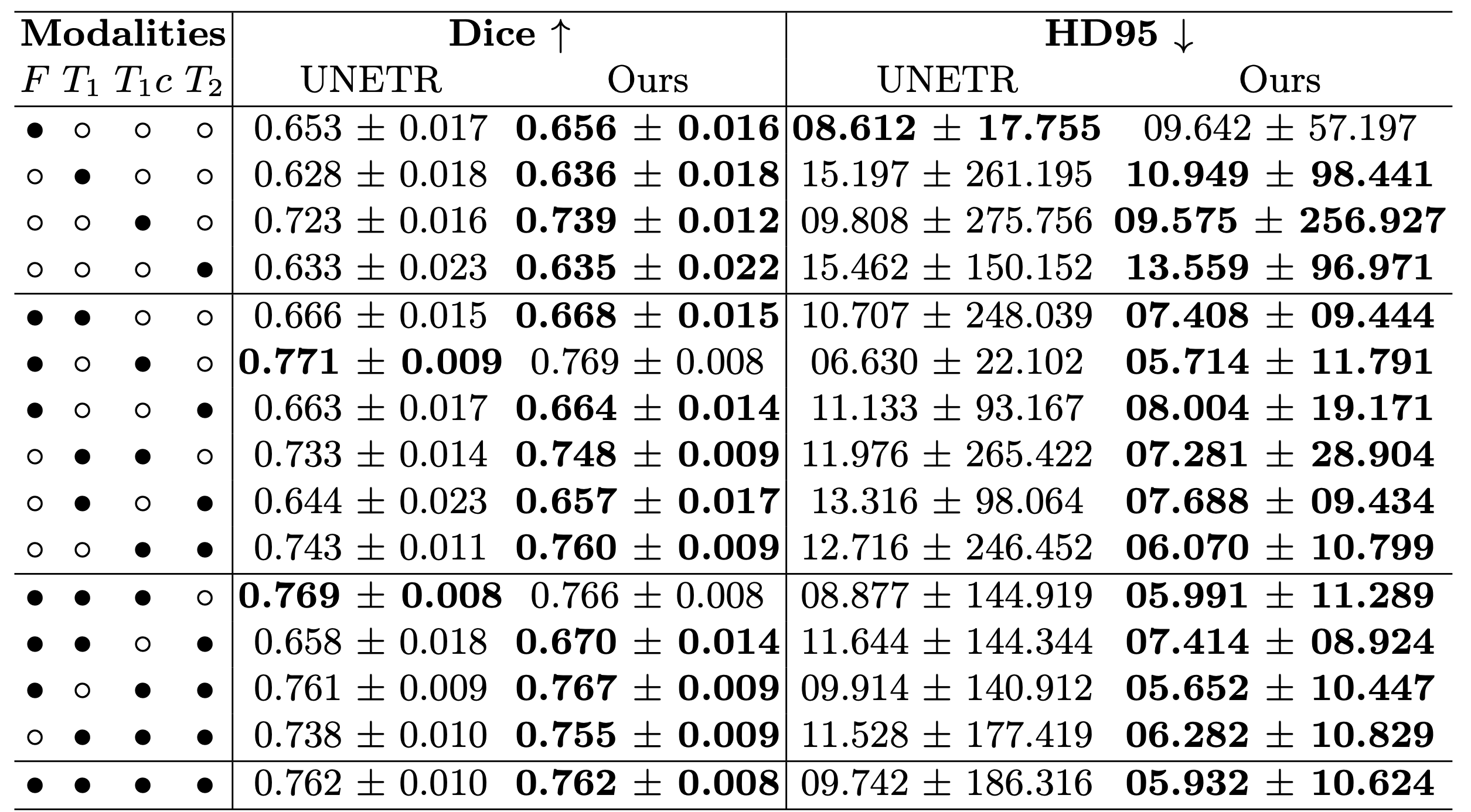}
    \caption{Dice and HD95 score details on MAG inputs of MSD dataset trained by MAG-MS compared with individually trained UNETR.}
    \label{table.msdUNETR}
\end{table}
\setlength{\textfloatsep}{6pt}

The efficiency of MAG-MS compared to individual UNETR models on the MSD dataset is also noteworthy. Each unique UNETR model required ~26 hours to train. Considering there are four modalities, training a model for each of the 15 combinations took ~390 hours. In contrast, MAG-MS only needed to be trained once, reducing overall training time by over 75\%. As the number of modalities increases, the possible input combinations increase exponentially, requiring more time for training individual models. The high efficiency of MAG-MS shows its great potential in practice where time and resources are limited. 

\section{Conclusion}
In this paper, we introduces the framework of MAG-MS and applies it to the investigation of input modalities effects in medical image segmentation. MAG-MS provides a versatile and efficient approach to handling partial available modalities in clinical practice. Our extensive experiments conducted on benchmark datasets demonstrated the high efficiency of MAG-MS and its superior segmentation performance than current state-of-the-art models, highlighting its strong potential for clinical applications. This works is not peer-reviewed.

\section{Disclosures}
Research reported in this publication was supported by the National Cancer Institute of the National Institutes of Health under Award Number T32CA009206, and the National Institute of Health under Award Number NIH R01HL153720. The content is solely the responsibility of the authors and does not necessarily represent the official views of the National Institutes of Health.

The authors acknowledge their collaboration with Modus Medical Devices and GE Healthcare in conducting this research.

\balance
\bibliographystyle{IEEEtranS}
\bibliography{references}

\begin{thebibliography}{10}
\providecommand{\url}[1]{#1}
\csname url@samestyle\endcsname
\providecommand{\newblock}{\relax}
\providecommand{\bibinfo}[2]{#2}
\providecommand{\BIBentrySTDinterwordspacing}{\spaceskip=0pt\relax}
\providecommand{\BIBentryALTinterwordstretchfactor}{4}
\providecommand{\BIBentryALTinterwordspacing}{\spaceskip=\fontdimen2\font plus
\BIBentryALTinterwordstretchfactor\fontdimen3\font minus
  \fontdimen4\font\relax}
\providecommand{\BIBforeignlanguage}[2]{{%
\expandafter\ifx\csname l@#1\endcsname\relax
\typeout{** WARNING: IEEEtranS.bst: No hyphenation pattern has been}%
\typeout{** loaded for the language `#1'. Using the pattern for}%
\typeout{** the default language instead.}%
\else
\language=\csname l@#1\endcsname
\fi
#2}}
\providecommand{\BIBdecl}{\relax}
\BIBdecl

\bibitem{BraTS-bestInputs}
\BIBentryALTinterwordspacing
S.~Bakas \emph{et~al.}, ``Identifying the best machine learning algorithms for
  brain tumor segmentation, progression assessment, and overall survival
  prediction in the brats challenge,'' 2018. [Online]. Available:
  \url{https://arxiv.org/abs/1811.02629}
\BIBentrySTDinterwordspacing

\bibitem{modalityVariantRec}
A.~Chartsias, T.~Joyce, M.~V. Giuffrida, and S.~A. Tsaftaris, ``Multimodal mr
  synthesis via modality-invariant latent representation,'' \emph{IEEE
  transactions on medical imaging}, vol.~37, no.~3, pp. 803--814, 2017.

\bibitem{monai}
M.~Consortium, ``Monai: Medical open network for ai,'' Jul 2022.

\bibitem{hyperDenseNet}
J.~Dolz \emph{et~al.}, ``Hyperdense-net: a hyper-densely connected cnn for
  multi-modal image segmentation,'' \emph{IEEE transactions on medical
  imaging}, vol.~38, no.~5, pp. 1116--1126, 2018.

\bibitem{handlingMissingSequences}
E.~Gr{\o}vik, D.~Yi, M.~Iv, E.~Tong, L.~B. Nilsen, A.~Latysheva, C.~Saxhaug,
  K.~D. Jacobsen, {\AA}.~Helland, K.~E. Emblem \emph{et~al.}, ``Handling
  missing mri sequences in deep learning segmentation of brain metastases: a
  multicenter study,'' \emph{NPJ digital medicine}, vol.~4, no.~1, p.~33, 2021.

\bibitem{multimodalMedicalImg}
Z.~Guo \emph{et~al.}, ``Deep learning-based image segmentation on multimodal
  medical imaging,'' \emph{IEEE Transactions on Radiation and Plasma Medical
  Sciences}, vol.~3, no.~2, pp. 162--169, 2019.

\bibitem{unetr}
A.~Hatamizadeh \emph{et~al.}, ``Unetr: Transformers for 3d medical image
  segmentation,'' in \emph{WCACV}, 2022, pp. 574--584.

\bibitem{hemis}
M.~Havaei \emph{et~al.}, ``Hemis: Hetero-modal image segmentation,'' in
  \emph{Medical Image Computing and Computer-Assisted Intervention--MICCAI
  2016: 19th International Conference, Athens, Greece, October 17-21, 2016,
  Proceedings, Part II 19}.\hskip 1em plus 0.5em minus 0.4em\relax Springer,
  2016, pp. 469--477.

\bibitem{knowledgeDistillation}
G.~Hinton, J.~Dean, and O.~Vinyals, ``Distilling the knowledge in a neural
  network,'' 03 2014, pp. 1--9.

\bibitem{mobilenets}
A.~G. Howard \emph{et~al.}, ``Mobilenets: Efficient convolutional neural
  networks for mobile vision applications,'' \emph{arXiv preprint
  arXiv:1704.04861}, 2017.

\bibitem{kdOnMultiModalAndMonoModal}
M.~Hu \emph{et~al.}, ``Knowledge distillation from multi-modal to mono-modal
  segmentation networks,'' in \emph{Medical Image Computing and Computer
  Assisted Intervention -- MICCAI 2020}.\hskip 1em plus 0.5em minus 0.4em\relax
  Cham: Springer International Publishing, 2020, pp. 772--781.

\bibitem{nnUNet}
F.~Isensee, P.~F. Jaeger, S.~A. Kohl, J.~Petersen, and K.~H. Maier-Hein,
  ``nnu-net: a self-configuring method for deep learning-based biomedical image
  segmentation,'' \emph{Nature methods}, vol.~18, no.~2, pp. 203--211, 2021.

\bibitem{fullyConnectedCRF}
K.~Kamnitsas \emph{et~al.}, ``Efficient multi-scale 3d cnn with fully connected
  crf for accurate brain lesion segmentation,'' \emph{Medical image analysis},
  vol.~36, pp. 61--78, 2017.

\bibitem{brats}
B.~H. Menze, A.~Jakab, S.~Bauer, J.~Kalpathy-Cramer, K.~Farahani, J.~Kirby,
  Y.~Burren, N.~Porz, J.~Slotboom, R.~Wiest \emph{et~al.}, ``The multimodal
  brain tumor image segmentation benchmark (brats),'' \emph{IEEE transactions
  on medical imaging}, vol.~34, no.~10, pp. 1993--2024, 2014.

\bibitem{fullyConvNets}
D.~Nie \emph{et~al.}, ``Fully convolutional networks for multi-modality
  isointense infant brain image segmentation,'' in \emph{ISBI}.\hskip 1em plus
  0.5em minus 0.4em\relax IEEE, 2016, pp. 1342--1345.

\bibitem{transfer-learning}
M.~Oquab \emph{et~al.}, ``Learning and transferring mid-level image
  representations using convolutional neural networks,'' \emph{cvpr}, 2014.

\bibitem{MR-only-review}
A.~M. Owrangi, P.~B. Greer, and C.~K. Glide-Hurst, ``Mri-only treatment
  planning: benefits and challenges,'' \emph{Physics in Medicine \& Biology},
  vol.~63, no.~5, p. 05TR01, 2018.

\bibitem{pytorch}
\BIBentryALTinterwordspacing
A.~Paszke \emph{et~al.}, ``Pytorch: An imperative style, high-performance deep
  learning library,'' in \emph{Advances in Neural Information Processing
  Systems 32}.\hskip 1em plus 0.5em minus 0.4em\relax Curran Associates, Inc.,
  2019, pp. 8024--8035. [Online]. Available:
  \url{http://papers.neurips.cc/paper/9015-pytorch-an-imperative-style-high-performance-deep-learning-library.pdf}
\BIBentrySTDinterwordspacing

\bibitem{msd}
A.~L. Simpson \emph{et~al.}, ``A large annotated medical image dataset for the
  development and evaluation of segmentation algorithms,'' \emph{arXiv
  preprint}, 2019.

\bibitem{diceOverlap}
C.~H. Sudre, W.~Li, T.~Vercauteren, S.~Ourselin, and M.~Jorge~Cardoso,
  ``Generalised dice overlap as a deep learning loss function for highly
  unbalanced segmentations,'' in \emph{Deep Learning in Medical Image Analysis
  and Multimodal Learning for Clinical Decision Support: Third International
  Workshop, DLMIA 2017, and 7th International Workshop, ML-CDS 2017, Held in
  Conjunction with MICCAI 2017, Qu{\'e}bec City, QC, Canada, September 14,
  Proceedings 3}.\hskip 1em plus 0.5em minus 0.4em\relax Springer, 2017, pp.
  240--248.

\bibitem{evaluation}
K.~A. Wahid \emph{et~al.}, ``Evaluation of deep learning-based multiparametric
  mri oropharyngeal primary tumor auto-segmentation and investigation of input
  channel effects: Results from a prospective imaging registry,''
  \emph{Clinical and translational radiation oncology}, vol.~32, pp. 6--14,
  2022.

\bibitem{iseg2017}
L.~Wang \emph{et~al.}, ``Benchmark on automatic six-month-old infant brain
  segmentation algorithms: the iseg-2017 challenge,'' \emph{IEEE transactions
  on medical imaging}, vol.~38, no.~9, pp. 2219--2230, 2019.

\bibitem{foundationsOfMCL}
A.~Zadeh, P.~P. Liang, and L.-P. Morency, ``Foundations of multimodal
  co-learning,'' \emph{Information Fusion}, vol.~64, pp. 188--193, 2020.

\bibitem{multistreamFCN}
G.~Zeng and G.~Zheng, ``Multi-stream 3d fcn with multi-scale deep supervision
  for multi-modality isointense infant brain mr image segmentation,'' in
  \emph{2018 IEEE 15th International Symposium on Biomedical Imaging (ISBI
  2018)}.\hskip 1em plus 0.5em minus 0.4em\relax IEEE, 2018, pp. 136--140.

\bibitem{paSeg}
S.~Zhai \emph{et~al.}, ``Pa-seg: Learning from point annotations for 3d medical
  image segmentation using contextual regularization and cross knowledge
  distillation,'' \emph{CoRR}, vol. abs/2208.05669, 2022.

\bibitem{selfDistillation}
L.~Zhang, J.~Song, A.~Gao, J.~Chen, C.~Bao, and K.~Ma, ``Be your own teacher:
  Improve the performance of convolutional neural networks via self
  distillation,'' in \emph{Proceedings of the IEEE/CVF International Conference
  on Computer Vision}, 2019, pp. 3713--3722.

\bibitem{diceLoss}
R.~Zhao \emph{et~al.}, ``Rethinking dice loss for medical image segmentation,''
  in \emph{ICDM}.\hskip 1em plus 0.5em minus 0.4em\relax IEEE, 2020, pp.
  851--860.

\end{thebibliography}

\clearpage

\section{Appendix}

\subsection{Missing proof}
\begin{proof}
The KL-Divergence, also known as relative entropy, between the fused output $\tilde{f}$ given the multi-modality input $x$ and the single-modality output $f_i$ given a single-modality input $x_i$ in our model is given by:
    \begin{equation}
        \begin{split}
        & D_{KL}(M\|S) = D_{KL}((\tilde{f}(x)|x)\|(f_i(x_i)|\{x_i; i \in S\})) \\
        & = - P(\tilde{f}(x)|x) log \frac{P(f_i(x_i)|\{x_i; i \in S\})}{P(\tilde{f}(x)|x}) \\
        & = - P(\tilde{f}(x)|x) log P(f_i(x_i)|\{x_i; i \in S\}) + P(\tilde{f}(x)|x) log P(\tilde{f}(x)|x)
        \end{split}
    \end{equation}
where $P(\cdot|\cdot)$ denotes the conditional probability distribution of the output given testing inputs. 

MAG learning relies on the combination of multiple modalities to achieve better model performance, and additional modalities always provide more information or better discrimination in a given task. So, it is reasonable to have
\begin{equation}
    \forall S \subset M; P(\tilde{f}(x)|x) > P(f_i(x_i)|\{x_i; i \in S\})
\end{equation}
Then, we have:
    \begin{equation}
        \begin{split}
        & D_{KL}(M\|S) \\
        & > - P(f_i(x_i)|\{x_i; i \in S\}) log P(f_i(x_i)|\{x_i; i \in S\}) + P(\tilde{f}(x)|x) log P(\tilde{f}(x)|x) \\
        & = H(f_i(x_i)|\{x_i; i \in S\}) - H(\tilde{f}(x)|x)
        \end{split}
    \end{equation}
    which can be rewritten as:
    \begin{equation}
        \label{eq.klDivRelation}
        H(f_i(x_i)|\{x_i; i \in S\}) <  H(\tilde{f}(x)|x) + D_{KL}(M\|S))
    \end{equation}
That is, $H(f_i(x_i)|\{x_i; i \in S\})$ is upper bounded by $H(\tilde{f}(x)|x ) + D_{KL}(M\|S))$.

Let $\Theta_{dec}$ be the shared parameters of the network decoder. During the training with all $M$ modalities, the prediction of the probabilistic model $P(\tilde{f}(x)|x)$ is maximized to obtain $\theta_M \in \Theta_{dec}$ and $\theta'_M \in \Theta_{dec}$ for training without and with self-distillation, respectively. Through multi-modality self-distillation, the KL-Divergence term between the fused modality output (teacher) and the single modality output (student) is minimized. So, $\theta'_M$ leads to:
\begin{equation}
\begin{split}
D_{KL}((\tilde{f}(x)|x)\|(f_i(x_i)|\{x_i; i \in S\}); \theta'_M) \\
<D_{KL}((\tilde{f}(x)|x)\|(f_i(x_i)|\{x_i; i \in S\}); \theta_M)
\end{split}
\end{equation}
Subsequently, $\theta'_M$ leads to a tighter upper bound for the conditional entropy $H(f_i(x_i)|\{x_i; i \in S\})$, and as a result, we have $H(f_i(x_i)|\{x_i; i \in S\}; \theta'_M) < H(f_i(x_i)|\{x_i; i \in S\}; \theta_M)$.
\qed
\end{proof}

\clearpage

\subsection{Score Details on Different Modality Combinations}
\vspace{-0.25in}

\begin{table}[h!]
    \centering
    \tiny
    \begin{tabular}{c c c c | c | c | c | c }
        \hline
        \multicolumn{4}{c|}{\textbf{Modalities}} & \multirow{2}{*}{\textbf{Mean}} & \multirow{2}{*}{\textbf{WT}} & \multirow{2}{*}{\textbf{ET}} & \multirow{2}{*}{\textbf{TC}} \\
        $F$ & $T_1$ & $T_{1}c$ & $T_2$ & & & & \\ 
        \hline
        $\bullet$ & $\circ$ & $\circ$ & $\circ$ & 0.656 $\pm$ 0.128 & 0.774 $\pm$ 0.090 & 0.511 $\pm$ 0.255 & 0.683 $\pm$ 0.245 \\
        $\circ$ & $\bullet$ & $\circ$ & $\circ$ & 0.636 $\pm$ 0.135 & 0.693 $\pm$ 0.116 & 0.533 $\pm$ 0.260 & 0.681 $\pm$ 0.244 \\
        $\circ$ & $\circ$ & $\bullet $ & $\circ$ & 0.739 $\pm$ 0.108 & 0.721 $\pm$ 0.127 & 0.647 $\pm$ 0.248 & 0.847 $\pm$ 0.106 \\
        $\circ$ & $\circ$ & $\circ$ & $\bullet$ & 0.635 $\pm$ 0.150 & 0.726 $\pm$ 0.127 & 0.499 $\pm$ 0.243 & 0.681 $\pm$ 0.251 \\
        \hline
        $\bullet$ & $\bullet$ & $\circ$ & $\circ$ & 0.668 $\pm$ 0.121 & 0.775 $\pm$ 0.082 & 0.528 $\pm$ 0.263 & 0.700 $\pm$ 0.234 \\
        $\bullet$ & $\circ$ & $\bullet$ & $\circ$ & 0.769 $\pm$ 0.091 & 0.796 $\pm$ 0.092 & 0.648 $\pm$ 0.243 & 0.863 $\pm$ 0.081 \\
        $\bullet$ & $\circ$ & $\circ$ & $\bullet$ & 0.664 $\pm$ 0.120 & 0.773 $\pm$ 0.091 & 0.518 $\pm$ 0.253 & 0.701 $\pm$ 0.236 \\
        $\circ$ & $\bullet$ & $\bullet$ & $\circ$ & 0.748 $\pm$ 0.094 & 0.734 $\pm$ 0.105 & 0.651 $\pm$ 0.246 & 0.858 $\pm$ 0.082 \\
        $\circ$ & $\bullet$ & $\circ$ & $\bullet$ & 0.657 $\pm$ 0.130 & 0.737 $\pm$ 0.112 & 0.530 $\pm$ 0.255 & 0.702 $\pm$ 0.237 \\
        $\circ$ & $\circ$ & $\bullet$ & $\bullet$ & 0.760 $\pm$ 0.095 & 0.763 $\pm$ 0.100 & 0.652 $\pm$ 0.253 & 0.865 $\pm$ 0.074 \\
        \hline
        $\bullet$ & $\bullet$ & $\bullet$ & $\circ$ & 0.766 $\pm$ 0.092 & 0.790 $\pm$ 0.085 & 0.647 $\pm$ 0.249 & 0.860 $\pm$ 0.079 \\
        $\bullet$ & $\bullet$ & $\circ$ & $\bullet$ & 0.670 $\pm$ 0.118 & 0.772 $\pm$ 0.090 & 0.531 $\pm$ 0.258 & 0.707 $\pm$ 0.230 \\
        $\bullet$ & $\circ$ & $\bullet$ & $\bullet$ & 0.767 $\pm$ 0.092 & 0.796 $\pm$ 0.086 & 0.644 $\pm$ 0.252 & 0.861 $\pm$ 0.079 \\
        $\circ$ & $\bullet$ & $\bullet$ & $\bullet$ & 0.755 $\pm$ 0.093 & 0.759 $\pm$ 0.095 & 0.646 $\pm$ 0.255 & 0.860 $\pm$ 0.073 \\ 
        \hline
        $\bullet$ & $\bullet$ & $\bullet$ & $\bullet$ & 0.762 $\pm$ 0.092 & 0.790 $\pm$ 0.084 & 0.640 $\pm$ 0.255 & 0.857 $\pm$ 0.076 \\
        \hline
    \end{tabular}
    \caption{Dice scores details on different modality combination of MSD dataset by MAG-MS for whole tumor (WT), enhancing tumor (ET), and tumor core (TC). A $\bullet$ indicates that the corresponding modalities are available during testing.}
    \label{table.msdHd95}
    \vspace{-0.4in}
    \end{table}

\begin{table}[h!]
    \centering
    \tiny
    \begin{tabular}{c c c c | c | c | c | c }
        \hline
        \multicolumn{4}{c|}{\textbf{Modalities}} & \multirow{2}{*}{\textbf{Mean}} & \multirow{2}{*}{\textbf{WT}} & \multirow{2}{*}{\textbf{ET}} & \multirow{2}{*}{\textbf{TC}} \\
        $F$ & $T_1$ & $T_{1}c$ & $T_2$ & & & & \\ 
        \hline
        $\bullet$ & $\circ$ & $\circ$ & $\circ$ & 09.642 $\pm$ 07.563 & 11.541 $\pm$ 19.366 & 09.028 $\pm$ 05.216 & 08.358 $\pm$ 08.069 \\
        $\circ$ & $\bullet$ & $\circ$ & $\circ$ & 10.949 $\pm$ 09.922 & 14.740 $\pm$ 19.783 & 11.825 $\pm$ 13.940 & 06.283 $\pm$ 04.570 \\
        $\circ$ & $\circ$ & $\bullet$ & $\circ$ & 09.575 $\pm$ 16.029 & 09.677 $\pm$ 11.578 & 11.270 $\pm$ 19.065 & 07.778 $\pm$ 20.595 \\
        $\circ$ & $\circ$ & $\circ$ & $\bullet$ & 13.559 $\pm$ 0.9.847 & 21.152 $\pm$ 21.924 & 11.529 $\pm$ 09.303 & 08.266 $\pm$ 10.001 \\
        \hline
        $\bullet$ & $\bullet$ & $\circ$ & $\circ$ & 07.408 $\pm$ 03.073 & 07.417 $\pm$ 08.614 & 08.372 $\pm$ 04.181 & 06.435 $\pm$ 04.876 \\
        $\bullet$ & $\circ$ & $\bullet$ & $\circ$ & 05.714 $\pm$ 03.434 & 07.256 $\pm$ 08.549 & 06.435 $\pm$ 03.744 & 03.451 $\pm$ 04.190 \\
        $\bullet$ & $\circ$ & $\circ$ & $\bullet$ & 08.004 $\pm$ 04.379 & 09.477 $\pm$ 11.591 & 07.915 $\pm$ 03.708 & 06.621 $\pm$ 05.327 \\
        $\circ$ & $\bullet$ & $\bullet$ & $\circ$ & 07.281 $\pm$ 05.376 & 10.836 $\pm$ 15.514 & 07.340 $\pm$ 04.918 & 03.669 $\pm$ 03.899 \\
        $\circ$ & $\bullet$ & $\circ$ & $\bullet$ & 07.688 $\pm$ 03.071 & 08.355 $\pm$ 08.433 & 08.690 $\pm$ 04.892 & 03.374 $\pm$ 03.617 \\
        $\circ$ & $\circ$ & $\bullet$ & $\bullet$ & 06.070 $\pm$ 03.286 & 08.001 $\pm$ 08.389 & 06.835 $\pm$ 04.735 & 03.734 $\pm$ 03.617 \\
        \hline
        $\bullet$ & $\bullet$ & $\bullet$ & $\circ$ & 05.991 $\pm$ 03.360 & 07.433 $\pm$ 08.538 & 06.969 $\pm$ 04.281 & 03.572 $\pm$ 03.979 \\
        $\bullet$ & $\bullet$ & $\circ$ & $\bullet$ & 07.414 $\pm$ 02.987 & 07.665 $\pm$ 08.464 & 08.510 $\pm$ 04.338 & 6.066 $\pm$ 04.580 \\
        $\bullet$ & $\circ$ & $\bullet$ & $\bullet$ & 05.652 $\pm$ 03.232 & 07.205 $\pm$ 08.538 & 06.253 $\pm$ 03.891 & 03.497 $\pm$ 03.807 \\
        $\circ$ & $\bullet$ & $\bullet$ & $\bullet$ & 06.282 $\pm$ 03.291 & 08.053 $\pm$ 08.363 & 07.161 $\pm$ 04.883 & 03.632 $\pm$ 03.884 \\ 
        \hline
        $\bullet$ & $\bullet$ & $\bullet$ & $\bullet$ & 05.932 $\pm$ 03.259 & 07.388 $\pm$ 08.478 & 06.813 $\pm$ 04.532 & 03.594 $\pm$ 03.809 \\
        \hline
    \end{tabular}
    \caption{HD95 scores details on different modality combination of MSD dataset by MAG-MS for whole tumor (WT), enhancing tumor (ET), and tumor core (TC). A $\bullet$ indicates that the corresponding modalities are available during testing.}
    \label{table.msdDice}
    \vspace{-0.4in}
\end{table}

\subsection{Additional Visualization Results of Prediction Masks by MAG-MS on Different Input Combinations}
\vspace{-0.3in}
\begin{figure}[h!]
    \centering
    \includegraphics[width=\textwidth]{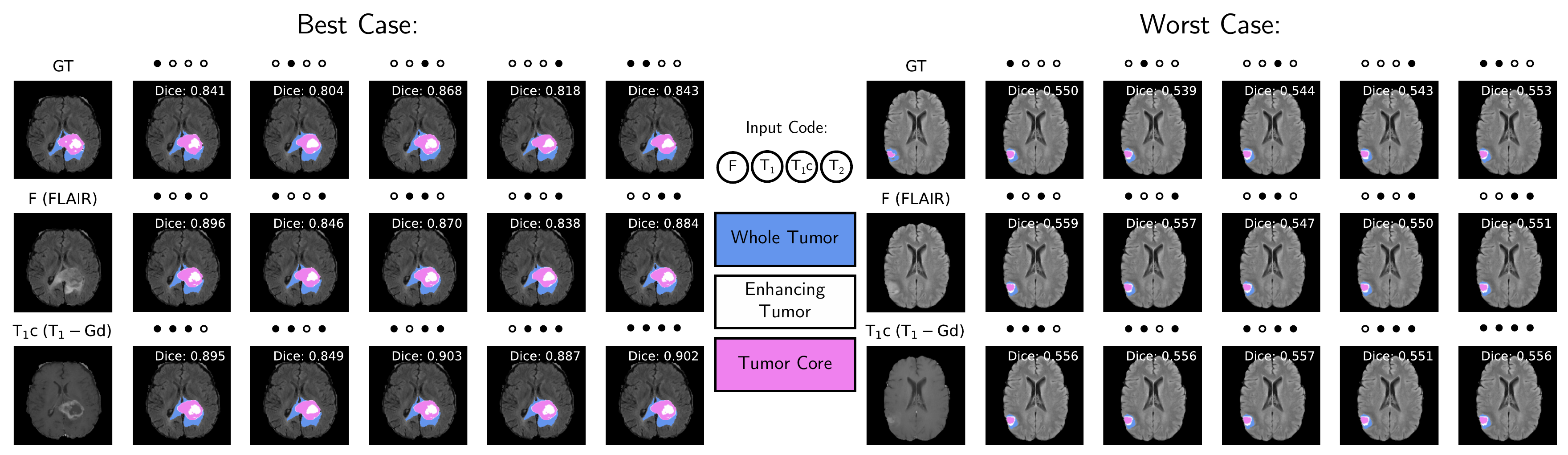}
    \caption{Best (left) and worst (right) prediction masks on different input combinations by our model. A $\bullet$ indicates that the corresponding modalities are available during testing.}
    \label{fig.predictions}
    \vspace{-0.2in}
\end{figure}

\end{document}